\documentclass[aps,pra,twocolumn,showpacs]{revtex4}

\usepackage{graphicx}
\usepackage{color}
\usepackage{bm}
\usepackage{amsmath, amsthm, amssymb,mathrsfs}
\usepackage{csquotes}
\usepackage[colorlinks=true, citecolor=blue,allcolors=blue]{hyperref}

\newcommand{\ME}[3]{ \mbox{$\langle #1\,|\,#2\,|\,#3\rangle $}}

\begin{document}

%\title{Effect of an optical laser in bichromatic ionization of Neon by XUV femtosecond pulse}
\title{Above-threshold ionization in neon produced by combining optical and bichromatic XUV femtosecond pulses}
%\title{Effect of an infrared laser on bichromatic ionization of Neon}

\author{Nicolas Douguet$^1$, Alexei~N.~Grum-Grzhimailo$^2$, and Klaus Bartschat$^1$}

\affiliation{$^1$Department of Physics and Astronomy, Drake University, Des Moines, Iowa 50311, USA}

\affiliation{$^2$Skobeltsyn Institute of Nuclear Physics, Lomonosov Moscow State University, 
Moscow 119991, Russia}

\date{\today}


\begin{abstract}
We consider the ionization of neon induced by a femtosecond laser pulse composed of overlapping, linearly polarized bichromatic 
extreme ultraviolet
and infrared fields. In particular, we study the effects of the infrared light on a two-pathway ionization 
scheme for which Ne $2s^22p^53s \, ^1P$ is used as intermediate state. 
Using time-dependent calculations, supported by a theoretical approach based on the strong-field approximation,
we analyze the ionization probability and the photo\-electron angular distributions associated with the different 
sidebands of the ionization spectrum. 
Complex oscillations of the angular distribution anisotropy parameters as a function of the infrared light 
intensity are revealed. Finally, we demonstrate that coherent control of the asymmetry is achievable by tuning 
the infrared frequency to a nearby electronic transition.  
\end{abstract}

\maketitle

\section{INTRODUCTION}
\label{sec:1}

The coherent control of quantum phenomena by light~\cite{Shapiro86,Brumer03} stands at 
the heart of future promising developments 
in a variety of scientific areas. Manipulating two-pathway quantum interferences in atomic 
ionization is one way to achieve 
coherent control of the photo\-electron angular distribution 
(PAD) (for example~\cite{Baranova92,Yin92,Schafer92,Ehlotzky01,Astapenko06}). 
The principle consists of ionizing an atomic system using the fundamental and second 
harmonic of a short laser pulse, 
thereby producing two distinct ionizing pathways characterized by one-photon 
and two-photon absorption. 
The latter process, referred to below as $\omega + 2\,\omega$, 
was recently studied experimentally in the extreme ultraviolet (XUV) regime at the free-electron
laser (FEL) FERMI in Trieste~\cite{Prince2016} using two color femto\-second (fs)
pulses for the ionization of neon. The efficiency of the two-photon ionization pathway was enhanced by choosing
one of the Ne $2p^54s$ states with total electronic angular momentum~$J=1$ as an inter\-mediate stepping stone.
Coherent control of the PAD asymmetry was achieved by varying the time delay, or the corresponding relative 
carrier envelope phase~(CEP), 
between the two harmonics, to an unprecedented precision of 3.1~attoseconds.
A description of the $\omega + 2\,\omega$ interference process in neon 
using $2p^53s$ as intermediate state is presented in Ref.~\cite{Douguet16b}.

\begin{figure}[h]
\includegraphics[width=8.5cm]{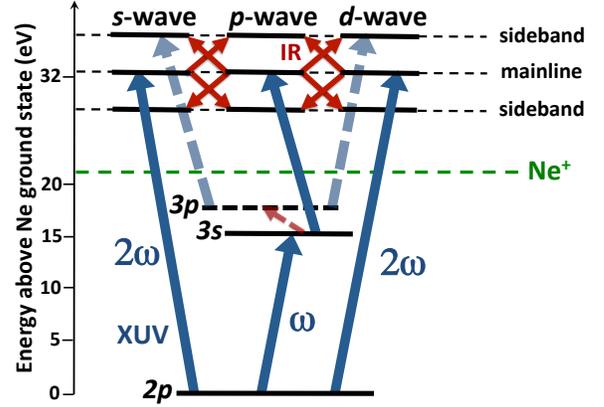}\\
  \caption{Scheme of the $\omega  + 2\,\omega + \Omega_0$ 
process in neon in the dipole approximation. The ionization is caused by the fundamental and second harmonic 
(solid blue arrows) of an XUV pulse whose fundamental frequency $\omega$ is tuned near the $2p\to3s$ transition.
The overlapping IR field induces ATI (small red arrows) transitions leading to sidebands in the spectrum. Only $s$, $p$,
and $d$-waves are displayed, although higher partial waves can contribute.
The dashed arrows represent additional paths created when the IR frequency is tuned near the $3s\to3p$ transition (see text).}
  \label{fig:1}
\end{figure}

In this paper we theoretically analyze the effect of an additional
comparatively weak infrared (IR) field (Keldysh parameter $\gamma \gg 1$)
on the $\omega + 2\,\omega$ ionization process and discuss 
the potential of the IR field to provide an additional degree of freedom to control the PAD. 

The presence of the IR field 
ultimately leads to the well-known phenomenon of above-threshold ionization 
(ATI)~\cite{Agostini79,Agostini81,Kruit81,Kruit83,Bucksbaum86}, resulting in sidebands in the photo\-electron spectrum
associated with the absorption or stimulated emission of one or several IR 
photons~\cite{Glover96,Paul01,Radcliffe07}. 
Many studies, both experimental and theoretical, of the sideband patterns in XUV + IR
ionization have been performed (for example~\cite{Okeefe04,Guyetand05,Meyer08,Meyer10,Richardson12,Maquet07}),
including PADs of the 
sidebands~\cite{Maquet07,Cionga93,Taieb00,Guyetand08,Haber09,Haber11,Doughty11,Sperl12,Dusterer13,Hutchinson13,Grum14,Dusterer16}. 
Recently such experiments with circularly polarized XUV beams from FELs became
feasible~\cite{Mazza16,Hartmann16}.

We chose neon as target for the study, because it is one of the atomic systems 
currently under investigation for coherent control experiments at the seeded FEL FERMI. 
We analyze ionization by a linearly polarized femto\-second pulse
whose electric field, taken along the $z$-axis, is given by 
$\bm{{\cal E}}(t) = \bm{{\cal E}}_{X}(t) + \bm{{\cal E}}_{IR}(t)$, where the XUV and IR components of the field are
\begin{eqnarray}
\label{eq:field-IR}
{\cal E}_{X}(t) &=& \bar{{\cal E}}_{X}f(t)[\cos(\omega t)+\eta_X \cos(2 \omega t + \varphi_X)] \,, \label{eq:1}\\
{\cal E}_{IR}(t) &=& \eta_0\bar{{\cal E}}_{X}f(t)\cos(\Omega_0 t + \varphi_0).
\end{eqnarray}

\noindent In the above equations, $f(t)$ is a smoothly varying pulse envelope, 
common to both XUV and IR fields, 
$\Omega_0$ is the infrared frequency, $\varphi_X$ and $\varphi_0$ are 
the CEPs of the second harmonic and the IR field, respectively, 
while the parameters $\eta_X$ and $\eta_0$ characterize their relative strength  
with respect to the fundamental amplitude $\bar{{\cal E}}_{X}$.
In our case the XUV pulse contains many optical cycles.  Hence
the CEP of the fundamental frequency $\omega$ in Eq.~(\ref{eq:field-IR}) is 
unimportant, and we set it to zero.  

\begin{figure}[b]
\includegraphics[width=8.5cm]{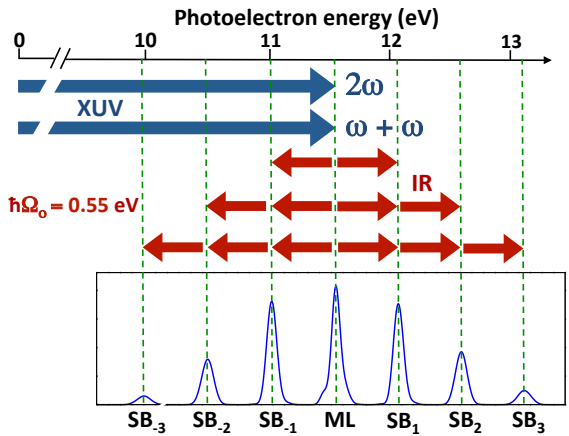}
  \caption{Example of a photo\-electron spectrum showing the different sidebands 
 associated with the minimum number of absorbed or emitted photons.}
  \label{fig:2}
\end{figure}

The ionization scheme of the $\omega + 2\,\omega + \Omega_0$ 
process is presented in Fig.~\ref{fig:1}.  We use the
single active electron (SAE) approximation to label neon electronic states, 
and the dipole approximation is employed throughout this study.
The scheme consists of tuning the fundamental 
frequency near the $2p^6 \, ^1S_0 \to 2p^53s \, ^1P_1$ excitation energy of neon, which
is associated with the one-electron transition $2p \to 3s$.
This results in a resonant two-photon ionization pathway,
which produces mostly $p$-wave photo\-electrons.
On the other hand, the second harmonic ionizes neon via non\-resonant one-photon absorption,
producing $s$-wave and $d$-wave photo\-electrons. 
These distinct pathways produce photo\-electrons
with partial waves of different parity, thereby leading to an asymmetric PAD.

%Of course, these interferences
%will have a noticeable effect only if the probability amplitudes squared of both paths are similar. 
%This condition can be fulfilled by adjusting adequately the ratio parameter $\eta_X$ between second harmonic 
%and fundamental field amplitudes in Eq. (\ref{eq:1}). 

The superimposed IR field creates equally spaced
sidebands around the mainline (ML).
This is illustrated in Fig.~\ref{fig:2} for an IR frequency $\Omega_0=0.55$~eV.
The sidebands are labeled as SB$_{\pm n}$ according to the minimum number of IR photons 
absorbed or emitted in the ATI process,
and $n>0$ is the band order. By definition, ML is the 0$^{th}$ order band.

Although only $s$, $p$, and $d$-waves are shown in Fig.~\ref{fig:1},
higher partial waves also contribute increasingly with higher IR intensity.
%$f$-waves also have a small 
%contribution at very low IR intensity. As the IR intensity increases, 
%the effects of higher partial waves become more important.
%Thus, we expect that the PAD at each sideband 
%will differ from the PAD at ML.
The differences in the relative contributions of partial waves lead 
to distinct PADs at the different sidebands.
Such dissimilarities may become
particularly pronounced in the extreme situation where the IR frequency is 
tuned to a nearby electronic state. 
Such a situation is shown in Fig.~\ref{fig:1} when the
IR frequency is set in resonance with the $3s\to3p$ one-electron transition, 
thus creating a pathway to ML and enhancing a \enquote{non-ATI} pathway for ionization
towards the lowest high-energy sideband. 
As will be shown below, this pathway can provide additional control on the PADs at different electron kinetic energies.

%As higher order terms are included in a time-dependent perturbative treatment, 
%the anisotropy parameters are first seen to behave quadratically 
%as a function of $\eta_0$ for $\eta_0\ll1$ and can 
%be further expanded in a power series of $\eta_0^2$ at higher intensity.

The weak field ($\eta_0\ll1$) characteristics of the different sidebands can be obtained 
from lowest-order perturbation theory~\cite{Lambropoulos76,Corrnier95}. Describing
the process at larger intensities requires further expansion of the ionization amplitude 
into a Born series including higher-order terms. 
However, such an expansion would necessitate the computation of free-free transition dipole moments, 
which are notoriously difficult to evaluate~\cite{Nikolopoulos06,Veniard90,Komninos12}. 
On the contrary, the Keldysh-Faisal-Reiss theory~\cite{Keldysh65,Faisal73,Reiss80,Reiss92} 
in the strong-field approximation (SFA) can be used to evaluate
characteristics of the ionization amplitude at the different sidebands, 
as recently demonstrated by Kazansky \textit{et al.}~\cite{Kazansky12}
and also described by Bauer~\cite{Bauer08,Bauer12} for strong-field photo\-ionization 
by a circularly polarized laser field. Therefore, we use the SFA to describe the ATI process
in the following development.

The next section provides a description of 
the numerical approach employed to solve 
the time-dependent Schr\"{o}dinger equation (TDSE)
and the SFA theoretical framework. In Sec.~\ref{sec:3}, we present our theoretical and numerical results 
and discuss the principal outcomes of the study. Section~\ref{sec:4} is devoted to our conclusions.
%Finally, an Appendix is provided with some further mathematical details. 

Unless otherwise indicated, atomic units are used throughout this manuscript.

\section{THEORY}
\label{sec:2}

Within the dipole approximation, the PAD is axially symmetric with respect to the
direction of linear polarization and is of the general form
\begin{equation}
\frac{dW}{d\Omega} = \frac{W_0}{4\pi}\left[1 + \sum^{\infty}_{k=1}\beta_kP_k (\cos\theta)\right],
\label{eq:ang_distribution}
\end{equation}
where $d\Omega$ represents the solid-angle element for a photo\-electron emitted into the direction defined by $(\theta, \phi)$, 
$W_0$ is the angle-integrated ionization probability, $P_k (\cos\theta)$ are Legendre polynomials, and $\beta_k$ are anisotropy parameters. 
Although not explicitly shown in Eq.~(\ref{eq:ang_distribution}), the angular distribution, ionization probability, 
and anisotropy parameters depend on the electron kinetic energy $\varepsilon$. 
One can obtain anisotropy parameters associated with a given band
by computing their averaged value ${\cal P}^{-1}\int \beta_kW_0d\varepsilon$
over the energy range spanned by the band, with its ionization probability given by ${\cal P} = \int W_0d\varepsilon$.

The left-right asymmetry is defined as
\begin{equation}
A(0^{\circ}) =\frac{W(0^\circ)-W(180^\circ)}{W(0^\circ)+W(180^\circ)},
\label{eq:asymmetry1}
\end{equation}
where $W(\theta)$ is the ionization signal in the $\theta$ direction.
The left-right asymmetry can readily be 
expressed in terms of the anisotropy parameters as~\cite{Grum15}
\begin{equation}
A(0^{\circ}) = \frac{\sum^{\infty}_{k=0}\beta_{2k+1}}{1+\sum^{\infty}_{k=1}\beta_{2k}}.
\label{eq:asymmetry2}
\end{equation}
One clearly sees that only the odd-rank anisotropy parameters are responsible for the asymmetry of the PAD.
In second-order time-dependent perturbation theory (PT), the PAD~(\ref{eq:ang_distribution})
is expressed as
\begin{equation}
\label{eq:PAD1}
\frac{dW}{d\Omega} = \sum_{M_f=0,\pm1}|U_{\bm{k},M_f}^{(1)}+U_{\bm{k},M_f}^{(2)}|^2,
\end{equation}
where we introduced the first- and second-order ionization amplitudes for a photo\-electron with 
asymptotic momentum~$\bm{k}$. 
In Eq.~(\ref{eq:PAD1}), the residual ion has orbital angular momentum $L_f=1$ and magnetic quantum number $M_f$.  Here we
limit our consideration to an atom with zero initial angular momentum. 

In the following we employ the approach of~\cite{Kazansky12,Kazansky14} to develop
the SFA for the $\omega + 2\,\omega + \Omega_0$ process, however, we use the Coulomb-Volkov (CV) approximation \cite{Jain78,Duchateau02, Rodriguez04,Arbo08}.
The first-order ionization amplitude for one XUV photon absorption with the electronic wave function
\enquote{dressed} in the IR field takes the form
\begin{equation}
U^{(1)}_{\bm{k},M_f} = -i\int_{0}^{t_M}{\cal E}_X(t)\ME{\Psi_f\psi^{CV}_{\bm{k}}}{\hat{d}_z}{\Psi_0} 
e^{-i\varepsilon_{2p}t}dt \,,
\label{eq:amplitude}
\end{equation}
where $\varepsilon_{2p}<0$ is the binding energy of the $2p$ electron and
\hbox{$t_M $} indicates the end of the pulse that started at $t=0$.
Furthermore, $\hat d_z=\sum_j z_j$ is the component of the electric dipole operator, where the sum is
taken over all atomic electrons, $\Psi_0$ 
is the neon ground state, $\Psi_f$ is the final ionic state of Ne$^+$,
and $\psi^{CV}_{\bm{k}}$ denotes the Coulomb-Volkov wave function \cite{Duchateau02}:
%~\cite{Volkov35}.
\begin{equation}
\psi^{CV}_{\bm{k}}(\bm r) = \varphi_{\bm{k}}^-(\bm r)\exp\left(i\bm{A}(t)\cdot\bm{r}-\frac{i}{2}\int_{\infty}^tdt'[\bm{k}+\bm{A}(t')]^2\right).
\label{eq:CV}
\end{equation}
In the above equation, $\bm{A}(t) = \int_{t}^\infty\bm{{\cal E}}_{IR}(t') \,dt'$ is 
the vector potential of the IR field while $\varphi_{\bm{k}}^-$ is an incoming
eigenstate of the field-free SAE Hamiltonian.
Since in the present situation $k\gg A(t)$, and $\bm{A}(t)\cdot\bm{r}\ll1$ over the extent of the $2p$ orbital of neon,
 we can approximate $\varphi_{\bm{k}}^-(\bm r)\exp\left(i\bm{A}(t)\cdot\bm{r}\right)\approx\varphi^-_{\bm{k}}(\bm r)$
in (\ref{eq:CV}) when evaluating (\ref{eq:amplitude}). Within the single-configuration model, the matrix element in~(\ref{eq:amplitude})
is reduced to a one-electron matrix element. 
Expanding the continuum function $\varphi^-_{\bm{k}}$ 
in partial waves and considering the electron initially in a $p$-orbital, the amplitude~(\ref{eq:amplitude}) can be cast into the form
%Extending Eqs.~(8) and~(9) of~\cite{Kazansky10} to the electron initially in a $p$-orbital,
%the amplitude~(\ref{eq:amplitude}) can be cast into the form
 \begin{eqnarray}
 &U^{(1)}_{\bm{k},M_f} = -i\left[{\cal A}^{(1)}_{\varepsilon s,m}{\cal F}_{s,m}(\bm{k}) +{\cal A}^{(1)}_{\varepsilon d,m}{\cal F}_{d,m}(\bm{k})\right],
 \label{eq:ionization_ampl}
\end{eqnarray}
where $m=-M_f$,
 \begin{eqnarray}
{\cal F}_{\ell,m}(\bm{k}) = \eta_X\bar{{\cal E}}_{X} \int_0^{t_M} \! dt \, f(t)Y_{\ell m}(\theta, \phi)\nonumber\\
~~~\times\exp\left[-i \! \int^{\infty}_t \!\!
dt'\left(\frac{1}{2}[\bm{k}+\bm{A}(t)]^2-(2\,\omega+\varepsilon_{2p})\right)\right] ,
\label{eq:ampl1}
\end{eqnarray}
$Y_{\ell m}(\theta, \phi)$ are spherical harmonics, and 
\begin{equation}
{\cal A}^{(1)}_{\varepsilon \ell,m} = i^{-\ell} e^{i\delta_{\ell}} \ME{\varepsilon \ell m}{z}{2p,m} \,.
\label{eq:Asm} 
\end{equation}
%\begin{eqnarray}
%{\cal A}^{(1)}_{\varepsilon s,m}&=&\delta_{m,0}\frac{e^{i\delta_s}}{3}d_{\varepsilon s,2p}\label{eq:Asm} \,, \\
%{\cal A}^{(1)}_{\varepsilon d,m}&=&i^{-2}\frac{e^{i\delta_d}}{\sqrt{15}} \, (1m,10|2m) \, 
%d_{\varepsilon d,2p}\label{eq:Adm} \,.
%\end{eqnarray}
Here we separated out the phase factor $i^{-\ell} e^{i\delta_{\ell}}$ from the continuum wavefunction with
$\delta_l$ as the scattering (potential plus Coulomb) phase, $\varepsilon=k^2/2$ is the asymptotic electron energy, and
$(\theta, \phi)$ are the detection angles associated with $\bm{k}$.

The integral in brackets of Eq.~(\ref{eq:ampl1}) can be evaluated analytically for
an infinitely long ``pulse'' (steady-state excitation), when $A(t) = -\left( \eta_0 \bar{{\cal E}}_{X}/\Omega_0 \right)
 \sin{(\Omega_0t + \varphi_0)}$. Neglecting the term proportional to $A^2$ in~(\ref{eq:ampl1}) and
using the Jacobi-Anger expansion, one obtains
\begin{eqnarray}
 {\cal F}_{\ell,m}(\bm{k}) & = & \eta_X\bar{{\cal E}}_{X} 
 \sum^{\infty}_{n=-\infty} i^nJ_n(q)  e^{-i (\varphi_X + n \varphi_0)} \nonumber\\
  & & ~~~\times \int_{0}^{t_M} \! dt \,f(t)Y_{\ell m}(\theta, \phi) e^{i (\varepsilon- \varepsilon_n)t} \,,
\label{eq:ampl3}
\end{eqnarray}
where $\varepsilon_n = 2\,\omega + \varepsilon_{2p} + n\Omega_0$,
$q= \left( \eta_0 kA /\Omega_0 \right) \cos\theta$, $J_n(q)$ is the $n^{\rm th}$ Bessel function of the first kind. 
Rearranging the terms in the sum we obtain (for $\ell=0$ and $\ell=2$)
\begin{equation}
{\cal F}_{\ell,m}(\bm{k}) = \sum^{\infty}_{n=-\infty}{\cal F}^{(n)}_{\ell,m}(\bm{k})T^{(1)}_n \,,
\label{eq:ampl4}
\end{equation}
with %%% (see Appendix)
\begin{equation} \label{eq:fn}
{\cal F}^{(n)}_{\ell,m}(\bm{k}) = i^n J_n(q) Y_{\ell m}(\theta, \phi) \,,
\end{equation}
and in the rotating-wave approximation,
\begin{equation}
\label{eq:T1}
T_n^{(1)} = \eta_X\bar{{\cal E}}_{X}e^{-i( \varphi_X + n \varphi_0)}\int_0^{t_M}f(t)
e^{i(\varepsilon-\varepsilon_n)t}dt \,.
\end{equation}

The process of  two-photon resonant absorption is domi\-nated
by the pathway $2p\to3s\to\varepsilon p$, thus producing a $p$-wave photo\-electron with $m=0$.
Therefore, we neglect the contribution of other intermediate states.
The second-order ionization amplitude is deduced from similar considerations 
 and takes the form
\begin{eqnarray} \label{eq:u2}
 U^{(2)}_{\bm{k},M_f} = i^2{\cal A}^{(2)}_{\varepsilon p,m}{\cal F}_{p,m}(\bm{k}),
\end{eqnarray}
where
\begin{eqnarray} \label{eq:ampl5}
{\cal F}_{p,m}(\bm{k})&=&\sum^{\infty}_{n=-\infty}{\cal F}^{(n)}_{p,m}(\bm{k})T^{(2)}_n,
\end{eqnarray}
%\begin{eqnarray}
with 
\begin{equation}
\label{eq:T2}
T_n^{(2)}  =  \bar{{\cal E}}_{X}^2 e^{-in \varphi_0} \! \int_0^{t_M} \!\! f(t) e^{i(\varepsilon-\varepsilon_n-\Delta \omega)t} 
\! \int^{t}_0 \! f(t') e^{i \Delta \omega t'}dt' ,
\end{equation}
where $\Delta \omega = \varepsilon_{3s}- \varepsilon_{2p} - \omega$ and
\begin{equation}
{\cal A}^{(2)}_{\varepsilon p,m} =  \delta_{m,0 } \, i e^{i\delta_p}
\ME{\varepsilon p,m}{z}{3s,m} \ME{3s,m}{z}{2p,m}.
\end{equation}
If the XUV pulse contains many cycles, the functions $T^{(1)}_n$ and $T^{(2)}_n$ will
be narrow peaks centered at the energy position $\varepsilon_n$ of SB$_n$, turning into
a \hbox{$\delta$-function} in the limit of continuous radiation. 
Consequently, one can describe the ionization at SB$_n$ by ignoring all terms but 
${\cal F}^{(n)}_{\ell,m}(\bm{k})$ in Eqs.~(\ref{eq:ampl4}) and~(\ref{eq:ampl5}).

Note that the CEP $\varphi_0$ is factored out as $\exp[in\varphi_0]$ in
the total photo\-ionization amplitude into the SB$_n$ sideband (see Eqs.~(\ref{eq:T1}) and (\ref{eq:T2})). 
Therefore, the observable quantities do not 
depend on the CEP of the overlapping IR pulse. This feature was confirmed 
to high accuracy by our numerical calculations. It is an important property since, 
in contrast to the relative CEP $\varphi_X$ between the XUV harmonics, 
which can be experimentally controlled to high precision~\cite{Prince2016}, 
$\varphi_0$ is hardly controllable and mostly chaotic. Hereafter, we set $\varphi_0 = 0$. 

Using Eq.~(\ref{eq:PAD1}) and collecting Eqs.~(\ref{eq:amplitude}),
(\ref{eq:ampl4}), (\ref{eq:fn}), (\ref{eq:u2}), and (\ref{eq:ampl5}), 
we obtain the angular distribution at SB$_n$ and ML in the form
\begin{eqnarray}
\frac{dW^{(n)}}{d\Omega} & = & J^2_n(q)\Big(\Big|{\cal A}^{(1)}_{\varepsilon s,0}Y_{00}(\theta,\phi)T^{(1)}_n\nonumber\\
 &  & + {\cal A}^{(1)}_{\varepsilon d,0}Y_{20}(\theta,\phi)T^{(1)}_n
-i{\cal A}^{(2)}_{\varepsilon p,0}Y_{10}(\theta,\phi)T^{(2)}_n\Big|^2\nonumber\\
 & & + 2\Big|{\cal A}^{(1)}_{\varepsilon d,1}Y_{21}(\theta,\phi)T^{(1)}_n\Big|^2\Big) \label{eq:ang_distr}  \\
 & \approx &   J^2_n(q) \frac{dW^{(0)}}{d\Omega} \label{eq:and1}
\end{eqnarray}
The form (\ref{eq:and1}) is typical for the SFA within the ``soft-photon approximation''~\cite{Maquet07}
(see also~\cite{Mazza16}).
The ionization probability and anisotropy parameters at SB$_n$ are, respectively, given by
\begin{eqnarray}
W^{(n)}&=&\int\frac{dW^{(n)}}{d\Omega}{d\Omega} \,, \label{eq:ionization}\\
\beta_k^{(n)}&=& \frac{2k+1}{W^{(n)}} \int\frac{dW^{(n)}}{d\Omega}P_k(\cos\theta)d\Omega.
\label{eq:beta_general}
\end{eqnarray}
In the limit $\eta_0\to0$, calculating $\beta_k^{(0)}$ in Eq.~(\ref{eq:beta_general}) and using $J_0(0)=1$,
one can recover the lowest-order PT expressions for the anisotropy parameters (see \hbox{Eqs.~(7)$-$(11)} of Ref.~\cite{Douguet16b}), 
when including only the contribution from the $3s$ intermediate state.

The elements ${\cal{A}}^{(i)}_{\ell,m}T_n^{(i)}$, $i=1,2$ in Eq.~(\ref{eq:ang_distr})
are extracted below from the ionization amplitude computed in the TDSE approach in the absence of the IR field. 
The latter elements carry information on the $\omega+2\,\omega$ process 
at ML for $\eta_0=0$, while the SFA predicts the effect of the IR field on
the characteristics of the different bands (ML + SB$_{\pm n}$) as a function of $\eta_0$.

%In the next section, we compare the results of this SFA approach with the numerical solutions 
%of the TDSE for the $\omega + 2\,\omega + \Omega_0$.

The general procedure to numerically solve the TDSE for linearly polarized electric field was 
described at length in Ref.~\cite{Grum06}
for the case of atomic hydrogen. For a multi-electron system, similar to~\cite{Douguet16b},
we solve the TDSE using the SAE approximation in 
an averaged electronic potential computed to reproduce as accurately as possible the energy levels of the neon excited states. 
The $2p\to3s$ excitation energy and the neon ionization energy obtained in our potential are, respectively, 16.36~eV~and 21.16~eV 
in the SAE approach, whereas the experimental values are 16.85~eV and 21.56~eV~\cite{Saloman04,NIST}.
The fine-structure splitting between the atomic states is not included in our non\-relativistic approach.
Since the $LS$-purity of the Ne$(2p^53s)^1P_1$ state is about 93\%~\cite{Machado82},
we neglect the $^3$P component of this configuration.  In this case, the SAE model
should work reasonably well.

In the SAE approximation, the one-electron wavefunction is initially 
propagated from the $(2p,m)$ orbital of neon.
At the end of the pulse, the wavefunction is projected onto continuum distorted-wave functions
of~Ne$^+$. All physical observables, such as the ionization probability and the PAD with its associated anisotropy
parameters, can then be computed in a straight\-forward way.
In order to represent an unpolarized atomic target, we propagated 
the three projections of the initial angular momentum~$m$ of the $2p$ orbital independently and subsequently average the results.
We included $\ell \le 14$ in order to ensure the numerical convergence of our predictions.

\section{Results and discussion}
\label{sec:3}

\begin{figure}[b]
\includegraphics[width=8.5cm]{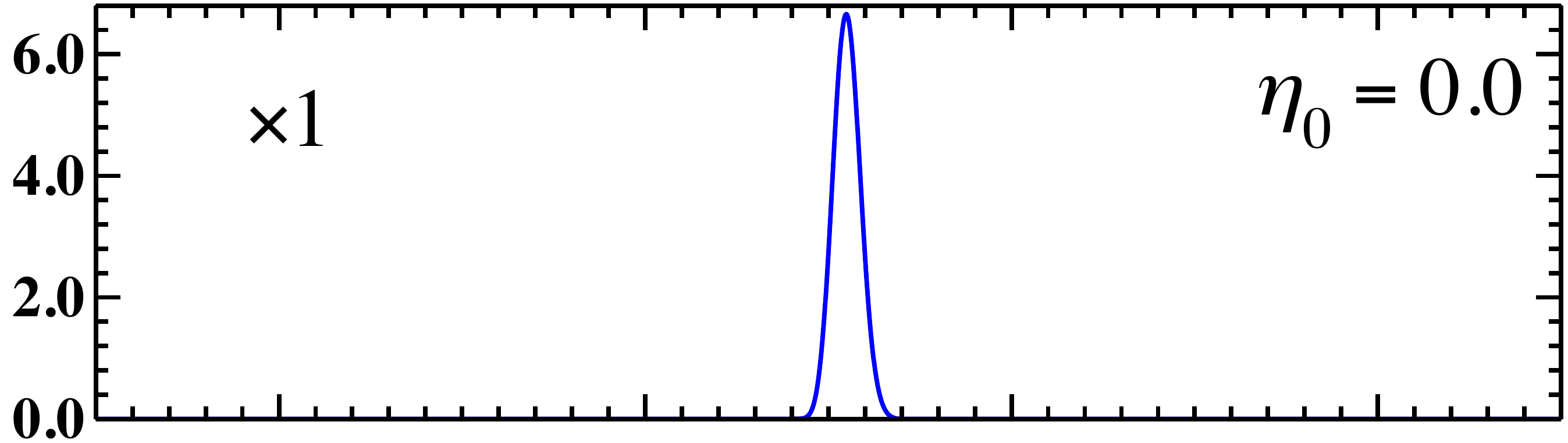}\\
\includegraphics[width=8.5cm]{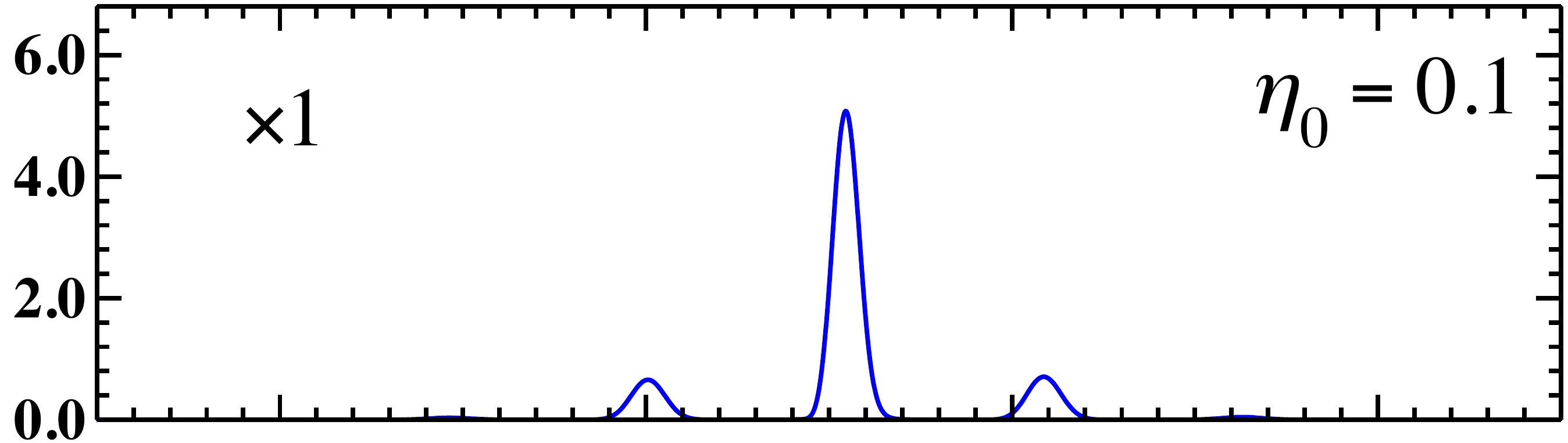}\\
\includegraphics[width=8.5cm]{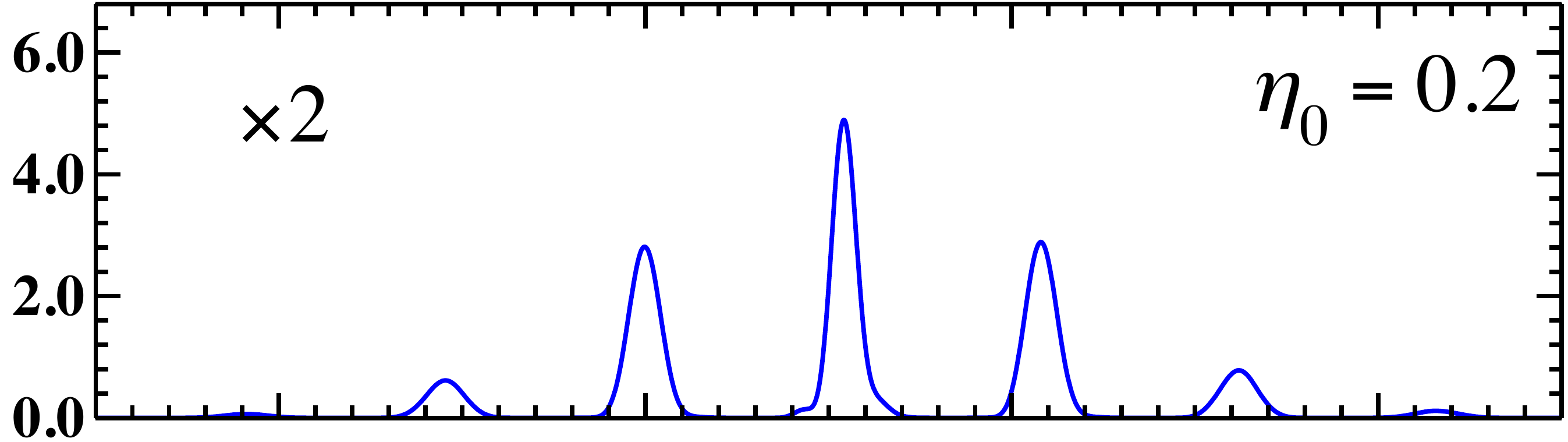}\\
\includegraphics[width=8.5cm]{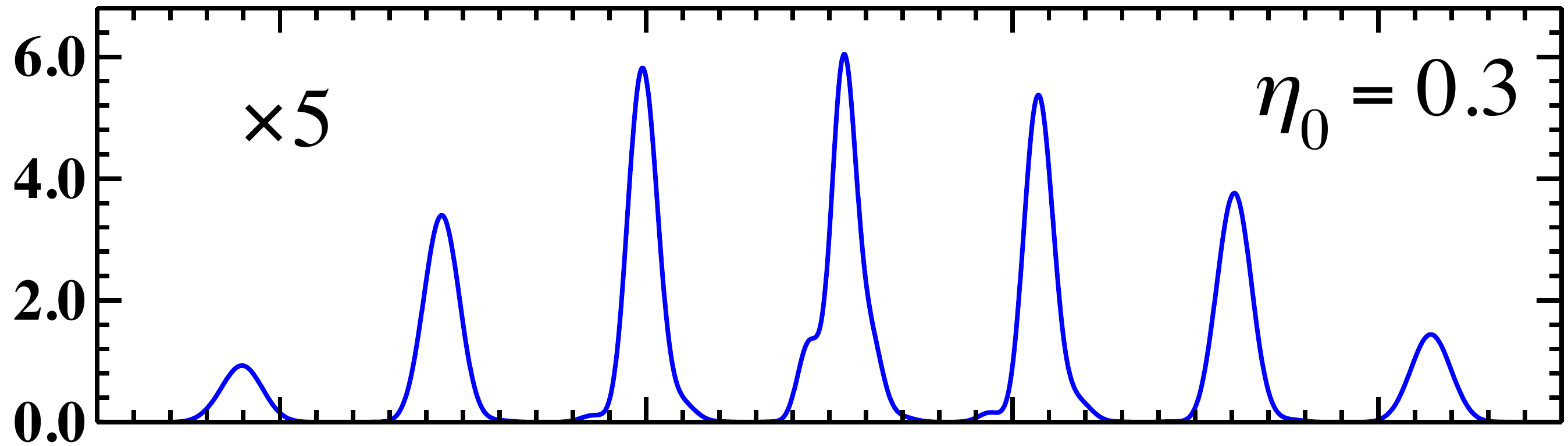}\\
\includegraphics[width=8.5cm]{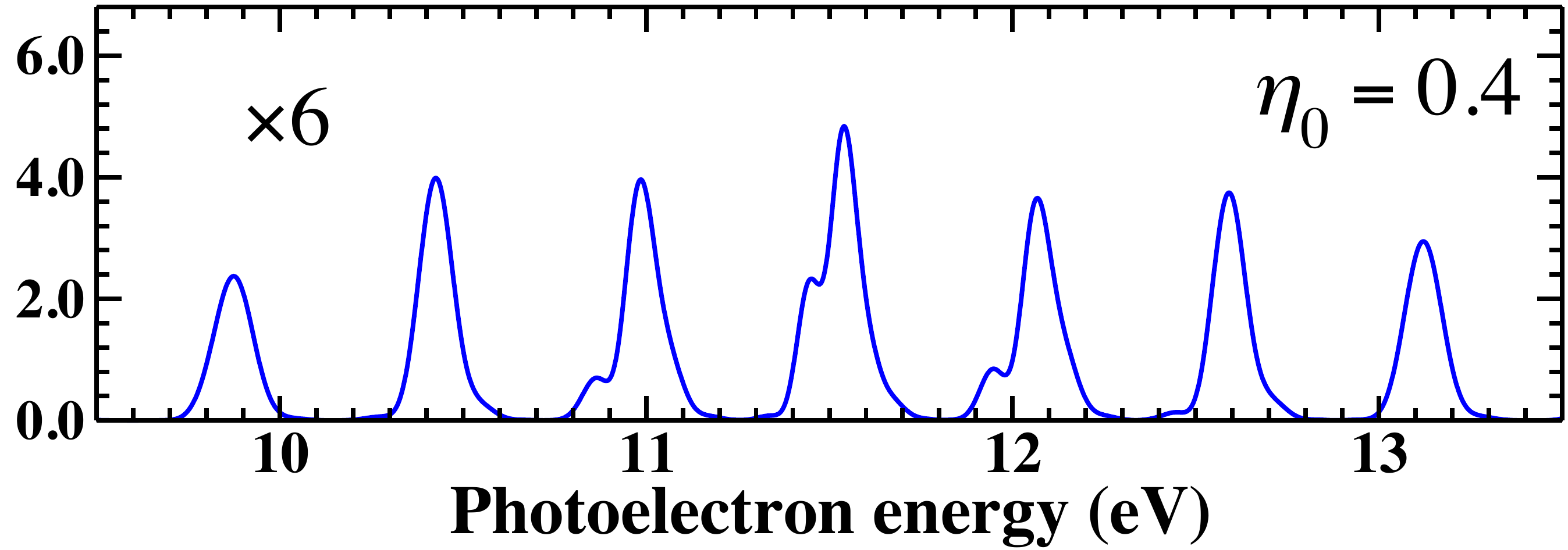}
  \caption{TDSE results for the ionization probability (in units of $10^{-3}$eV$^{-1}$) as a function of the photo\-electron energy for different 
  IR field amplitude ratios~$\eta_0$. The fundamental XUV intensity is $I = 10^{12}\rm{W/cm}^2$, $\eta_X = 0.1$, 
  and the infrared frequency is $\Omega_0 = \rm{0.55~eV}$.}
  \label{fig:3}
\end{figure}

The numerical simulations in this work were performed using pulse envelopes of the form
\hbox{$f(t) = \sin^2\Omega t$ ($\Omega=\omega/2N, 0\le t\le t_M$)}, where~$N$
is the number of XUV cycles. Hereafter, we take a pulse with $N=300$ cycles, 
corresponding to a FWHM of the intensity $\sim$ 27~fs. The amplitude 
ratio $\eta_X=0.1$ is fixed at a value producing a significant 
$\omega+2\,\omega$ interference. 
The intensity of the fundamental is chosen 
relatively low, 10$^{12}$W/cm$^2$, to ensure the applicability of the PT 
approach in describing the $\omega+2\,\omega$ process.

In order to minimize processes involving absorption or emission of IR photons prior to ionization, 
we choose a low IR frequency, $\Omega_0 = 0.55$~eV (corresponding to \hbox{$\approx 2.25\,\mu$m}) in the mid-infrared range. The IR field 
spans $N_0=10$ cycles in order to completely overlap with the XUV pulse. 
For such a low IR frequency, pathways of the form $\hbar\omega \pm \hbar\Omega_0 + \hbar\omega$ 
have a negligible effect, since no intermediate bound states are reachable 
by less than the absorption or emission of four IR photons. Multi\-photon ionization of the form 
$\hbar\omega + n\hbar\Omega_0$ is also an inefficient  process at such low IR intensity, because
the system should absorb $n\ge9$ photons to ionize from the $3s$ state.

\begin{figure}[htbp]
\includegraphics[width=8.cm]{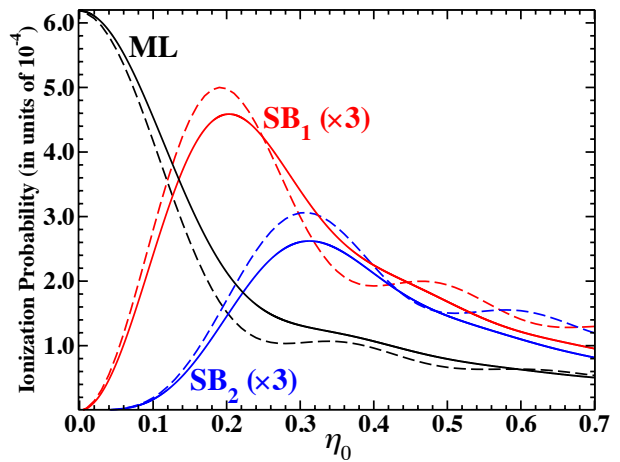}
\caption{Ionization probability associated with the different bands 
calculated in the TDSE (solid lines) and the SFA (dashed lines) approaches
as a function of the IR field amplitude ratio~$\eta_0$. }  
  \label{fig:4}
\end{figure}

\begin{figure*}[htbp]
 \begin{center}$
  \begin{array}{ccc}
  \vspace{-0.55cm}
\includegraphics[width=5.5cm]{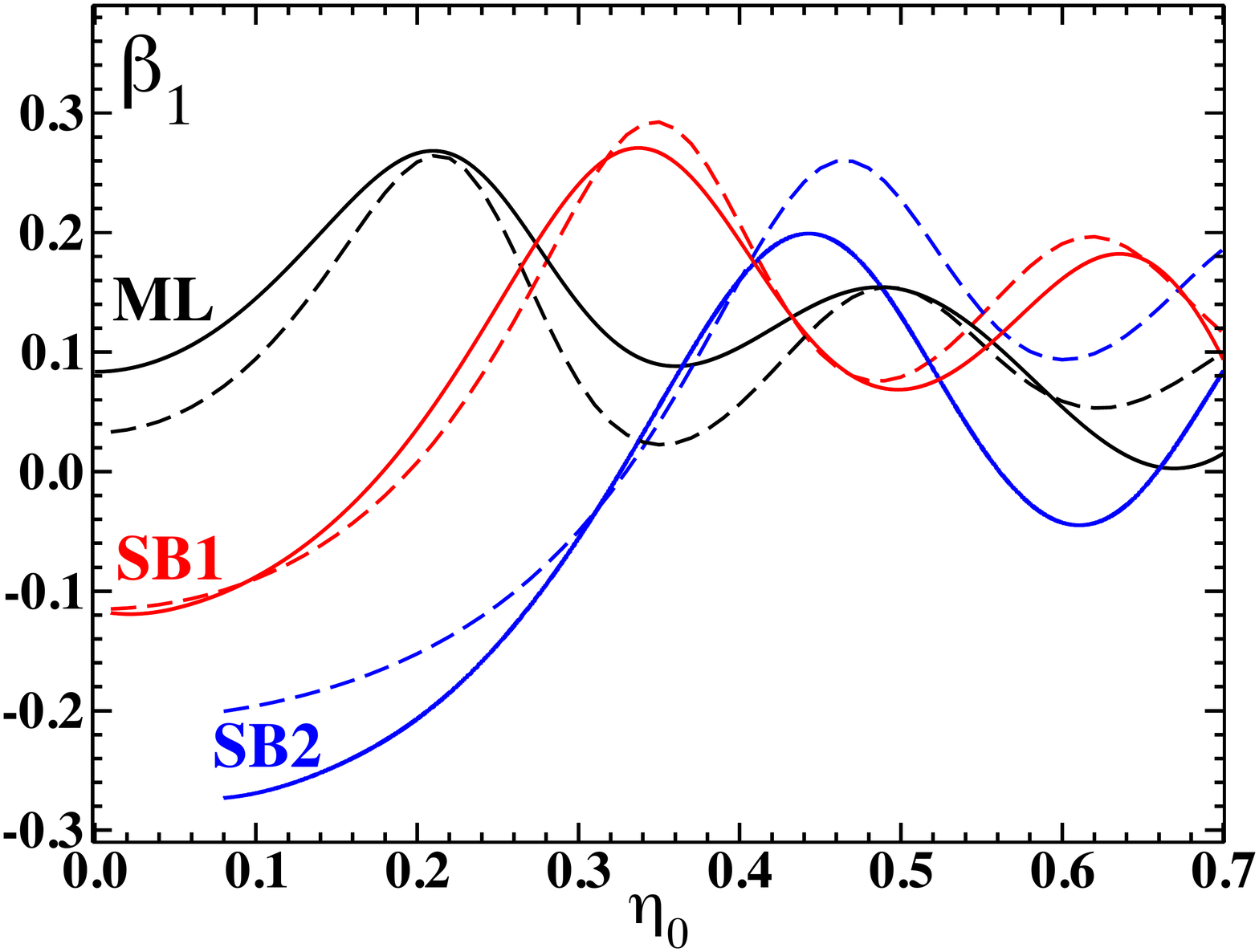} & \includegraphics[width=5.5cm]{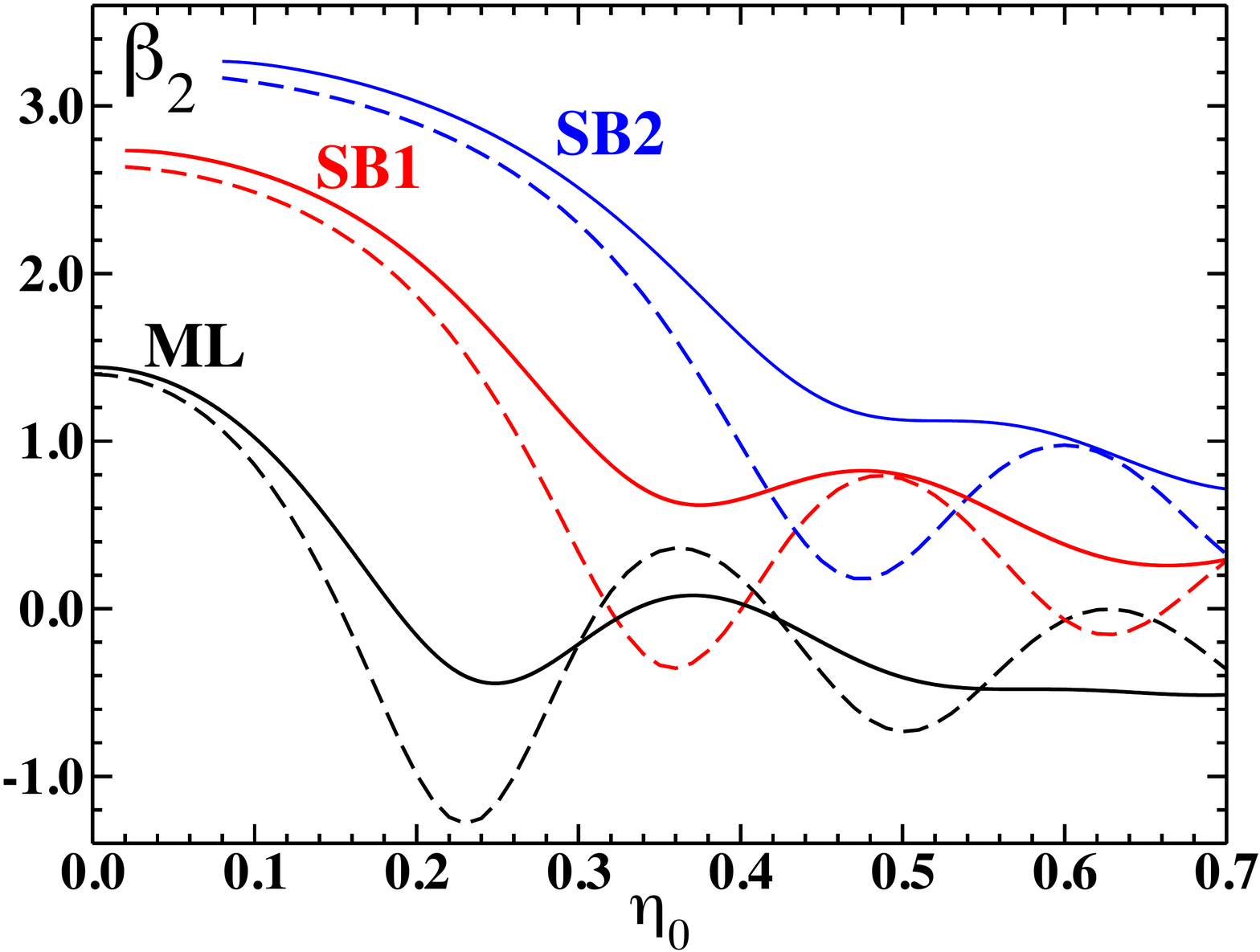}&\includegraphics[width=5.5cm]{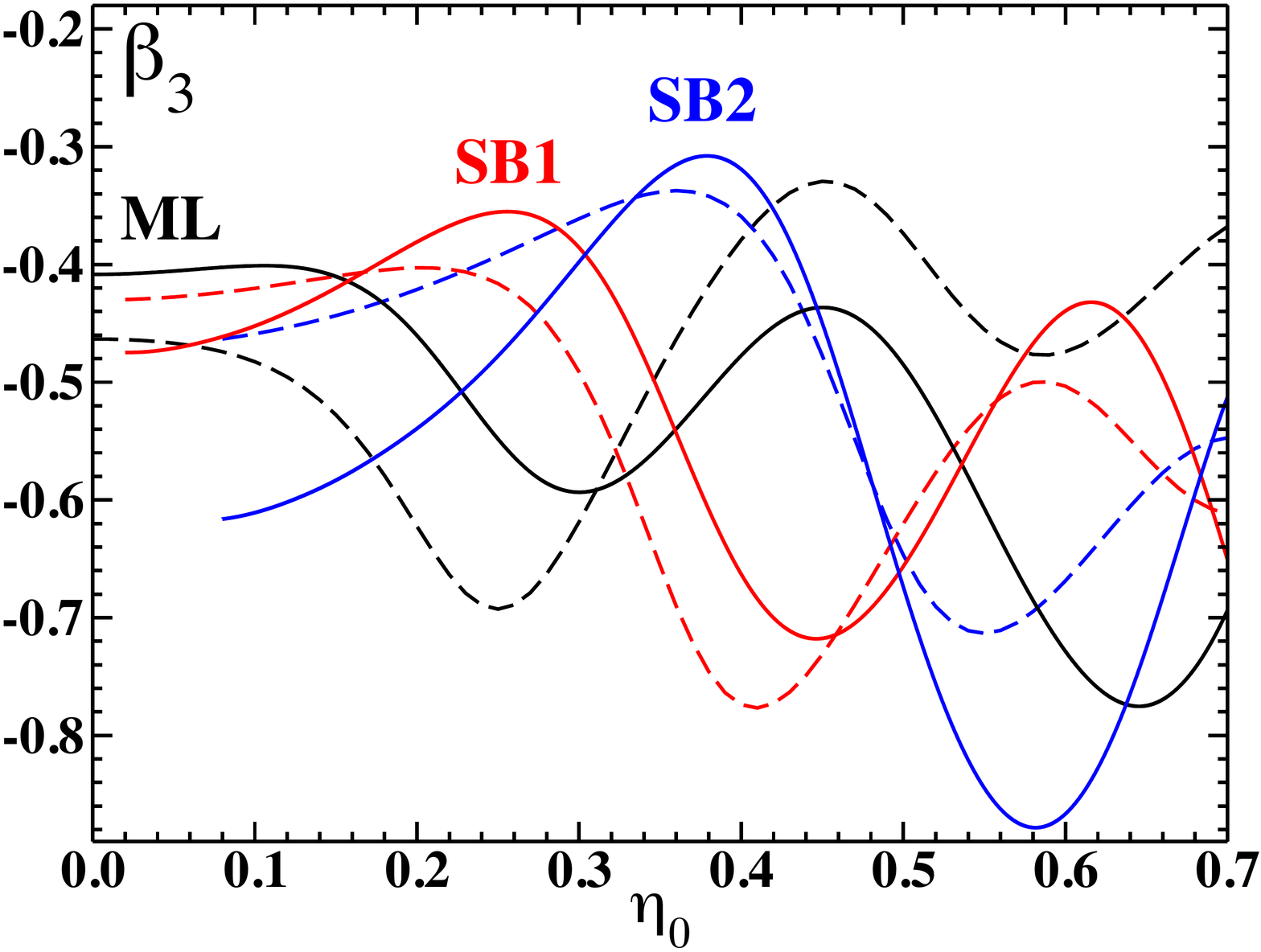}\\
\includegraphics[width=5.5cm]{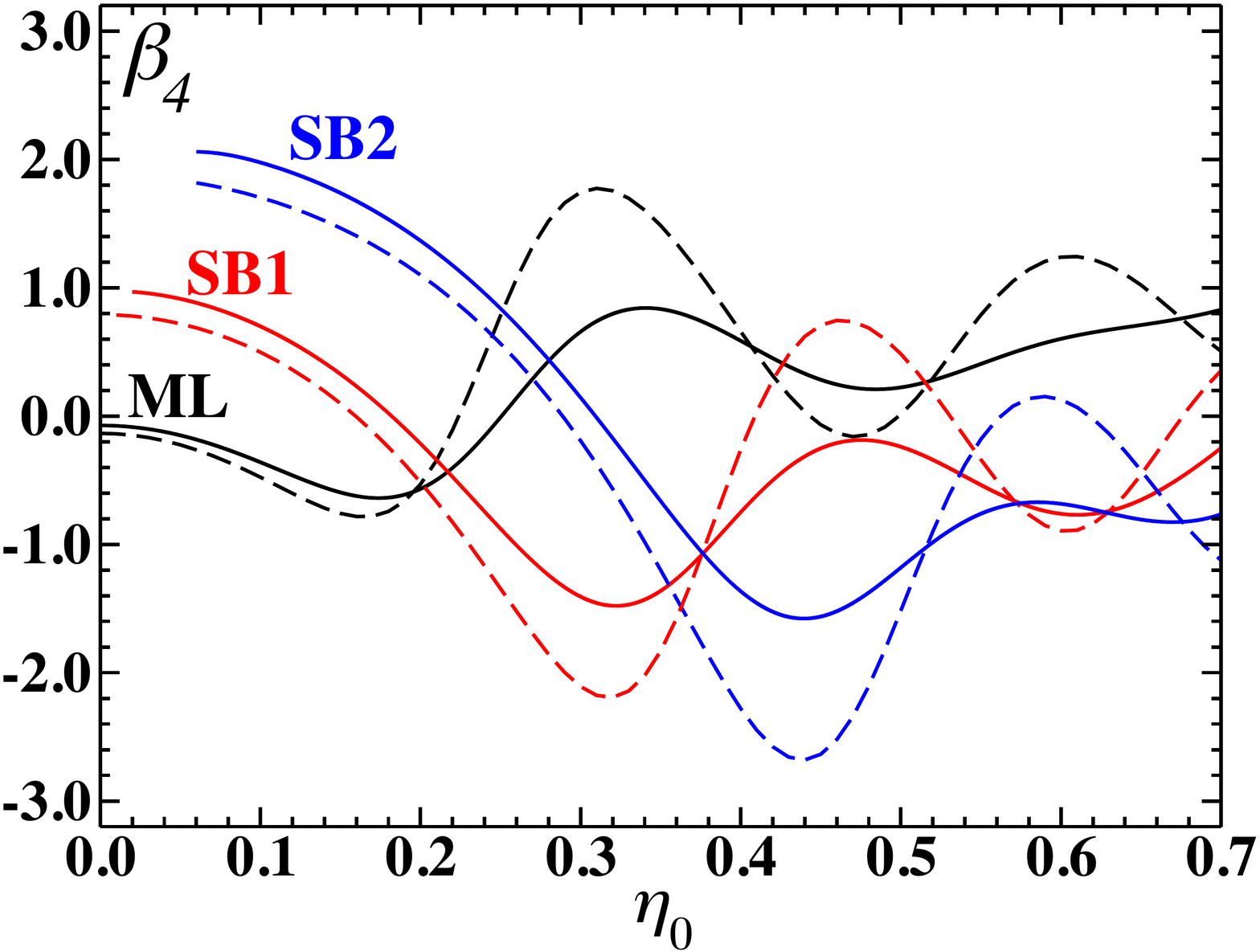} & \includegraphics[width=5.5cm]{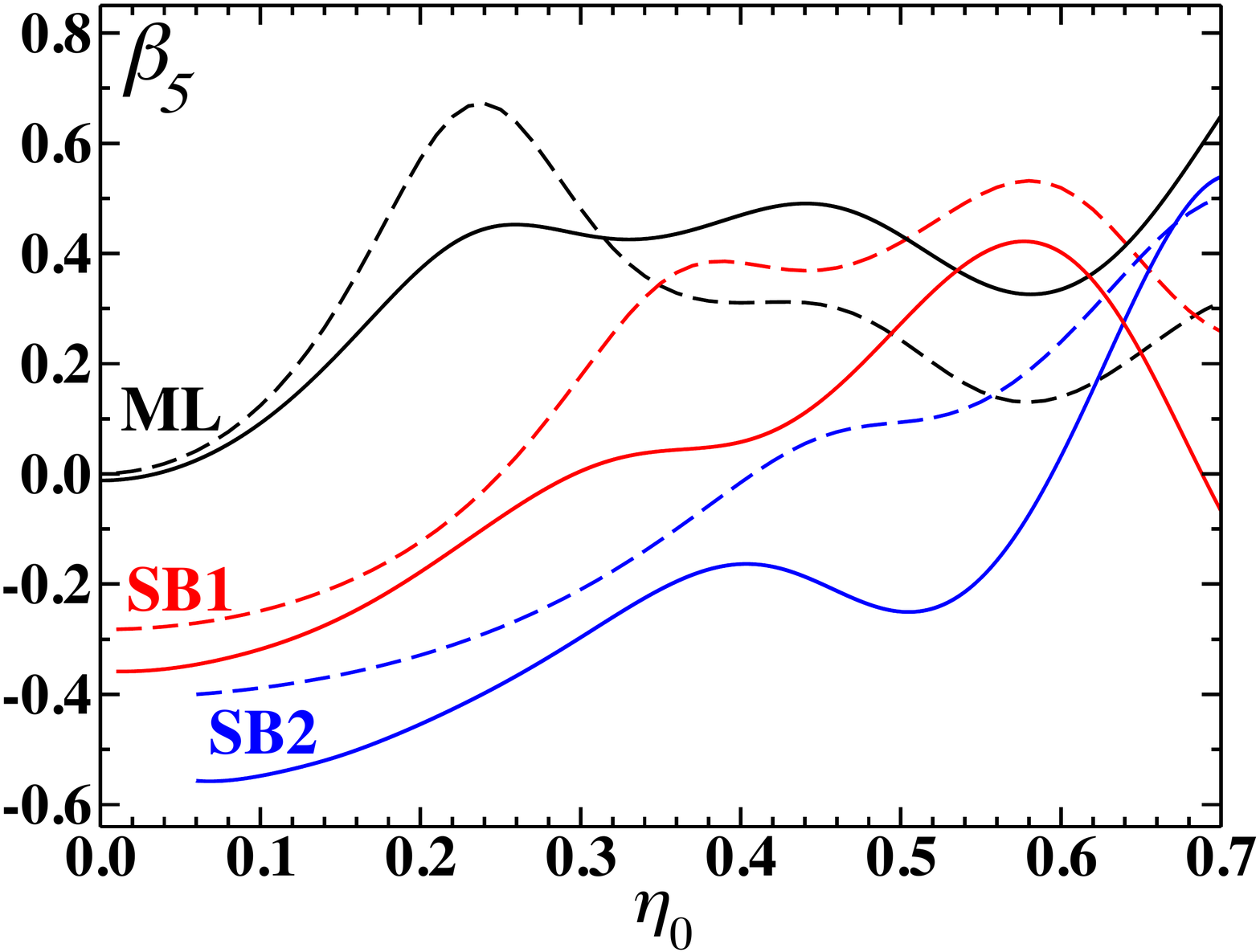}&\includegraphics[width=5.5cm]{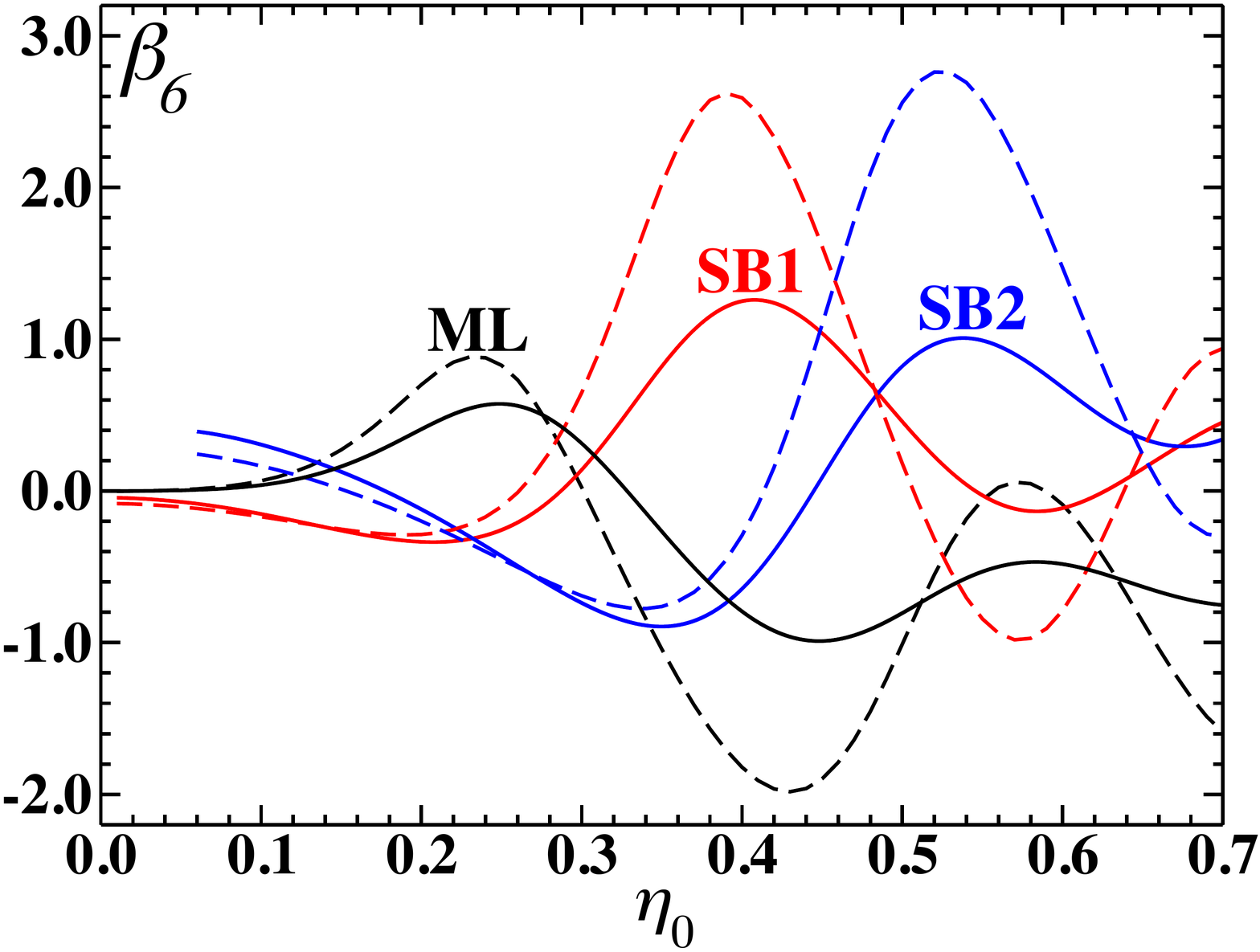}
\end{array}$
   \end{center}
 \caption{Anisotropy parameters $\beta_k$ ($1\le k\le6$), for $\varphi_X=0$, as a function of the IR field amplitude ratio $\eta_0$.
The labels are the same as in Fig.~\ref{fig:4}.}  
  \label{fig:5}
\end{figure*}

\subsection{General Appearance of Sidebands}

The angle-integrated photo\-electron spectra computed in the TDSE approach
are shown in Fig.~\ref{fig:3} at different amplitude ratios $\eta_0$.
Without the IR field, only ML appears in the spectrum. 
As the IR intensity is increased, SB$_{\pm1}$ are first created as a result of a single IR photon emission or absorption. 
Turning to larger intensities, SB$_{\pm n}$ with $n\ge2$ appear as a consequence of multiphoton processes.
The general tendencies for variations of the spectrum in the XUV + IR ionization
are well established and
are directly applicable in our case of $\omega + 2\omega +n \Omega_0$ ionization,
since the interference between the amplitudes of 
even-photon and odd-photon ionization vanishes in the angle-integrated photo\-electron spectra.
The ionization probability redistribution, and its dependence as a function of the IR field intensity, 
can be successfully modeled by the SFA approach~\cite{Maquet07}.
For $\eta_0\ge0.3$, nearby sidebands acquire an additional structure, 
which actually becomes a \enquote{double-peak} at higher intensities.  Its origin can be attributed 
to the ponderomotive energy and AC Stark shifts~\cite{Cormier01,Xu94,Moody77,Kruit83a}.

In the present study, we focus on the characteristics of ML and SB$_{n}$ for $|n|\le2$.
Since SB$_{n}$ and SB$_{-n}$ are formed, for small $n$, from approximately the same transition matrix elements, 
they exhibit similar characteristics within negligible differences. 
Thus, we will only show results for ML, SB$_1$ and SB$_2$ in the following development.
Similar results would be found for SB$_{-1}$ and~SB$_{-2}$, respectively.

The ionization probability associated with the different bands is presented in Fig.~\ref{fig:4} as a 
function of~$\eta_0$.
Since at high IR intensity the spectrum becomes strongly distorted, we only show the 
results for~$\eta_0\leq0.7$. 
The TDSE predictions are compared with the SFA results obtained using Eqs.~(\ref{eq:ang_distr}) and~(\ref{eq:ionization}).
Overall good agreement is obtained between the two sets of results. Ionization at ML
decreases monotonically with growing~$\eta_0$, whereas the weak-field ionization of SB$_1$ 
and SB$_2$ increases as $\eta_0^2$ and $\eta_0^4$,
respectively, in accordance with lowest-order (non\-vanishing) PT predicting an $\eta_0^{2n}$ dependence.
However, as the IR intensity is increased, the validity of PT breaks down rapidly.
The ionization probability in the sidebands reaches 
a maximum and gradually decreases as a function of intensity. The positions of the maxima 
of SB$_1$ and SB$_2$ at $\eta_0 \approx 0.2$ and $\eta_0 \approx 0.3$, respectively, 
are similar in the TDSE and SFA results. On the other hand, oscillations predicted by the SFA
are hardly apparent in the TDSE calculations. 

\begin{figure*}[htbp]
 \begin{center}$
  \begin{array}{ccc}
\includegraphics[width=5.5cm]{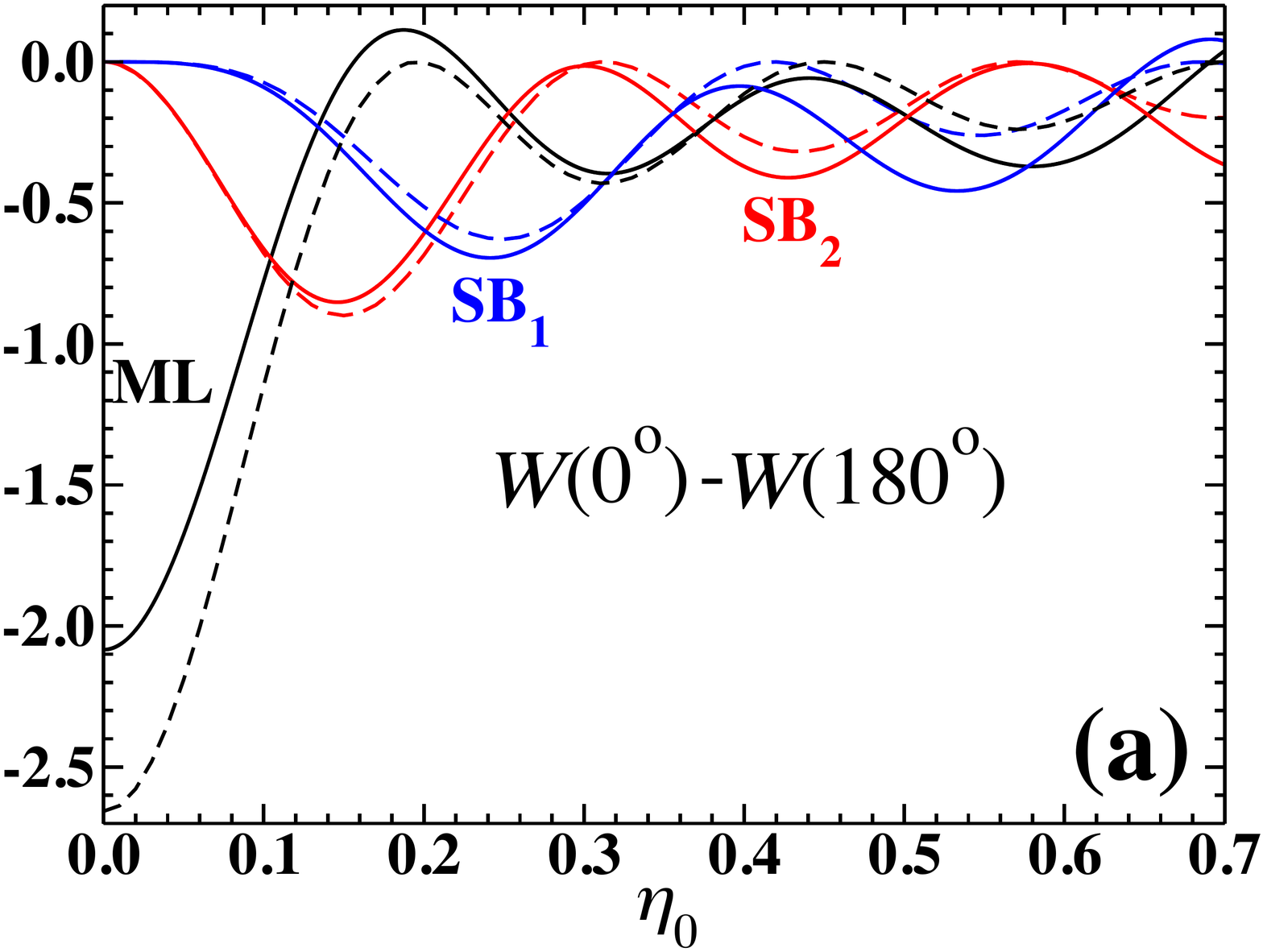} & \includegraphics[width=5.5cm]{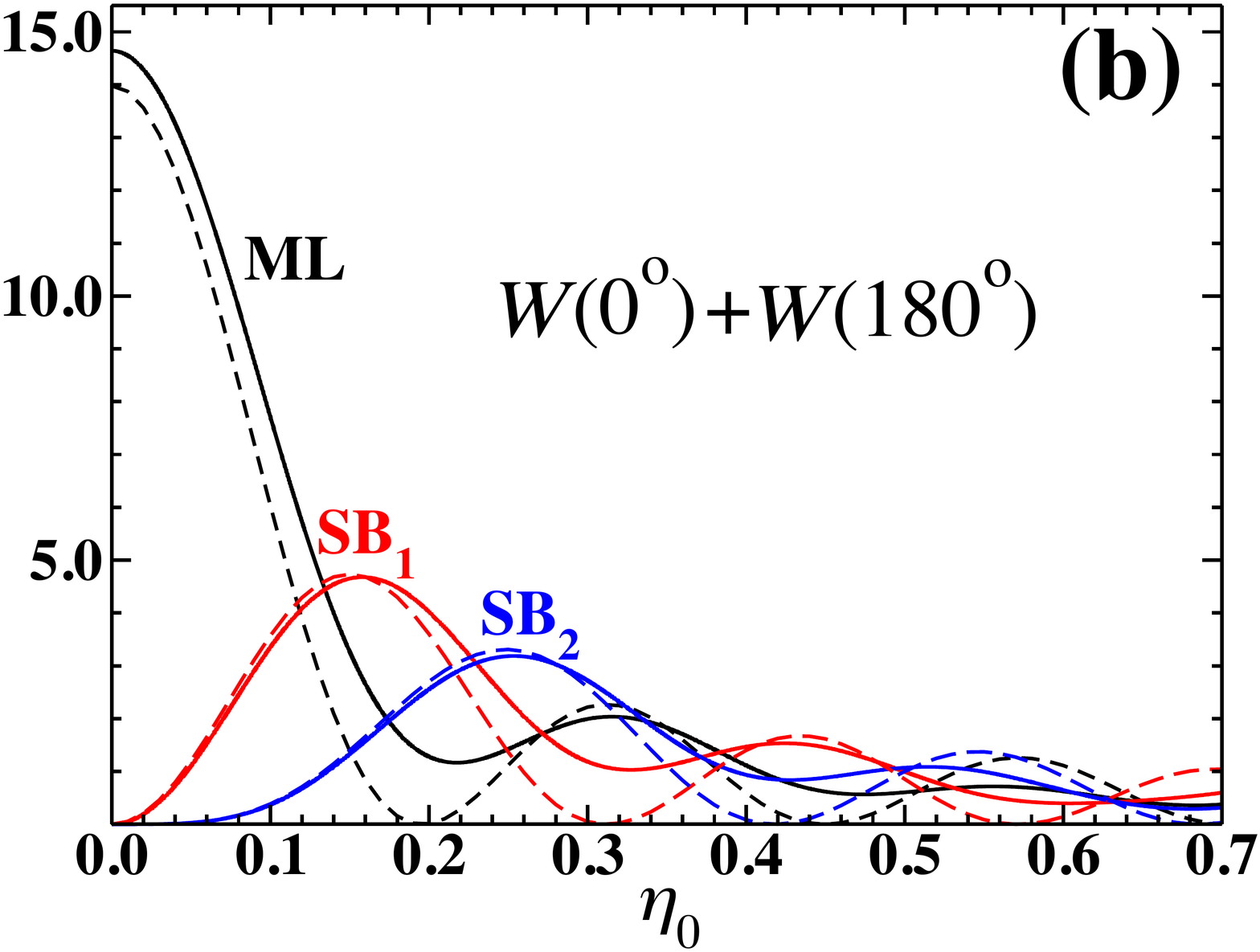}&\includegraphics[width=5.5cm]{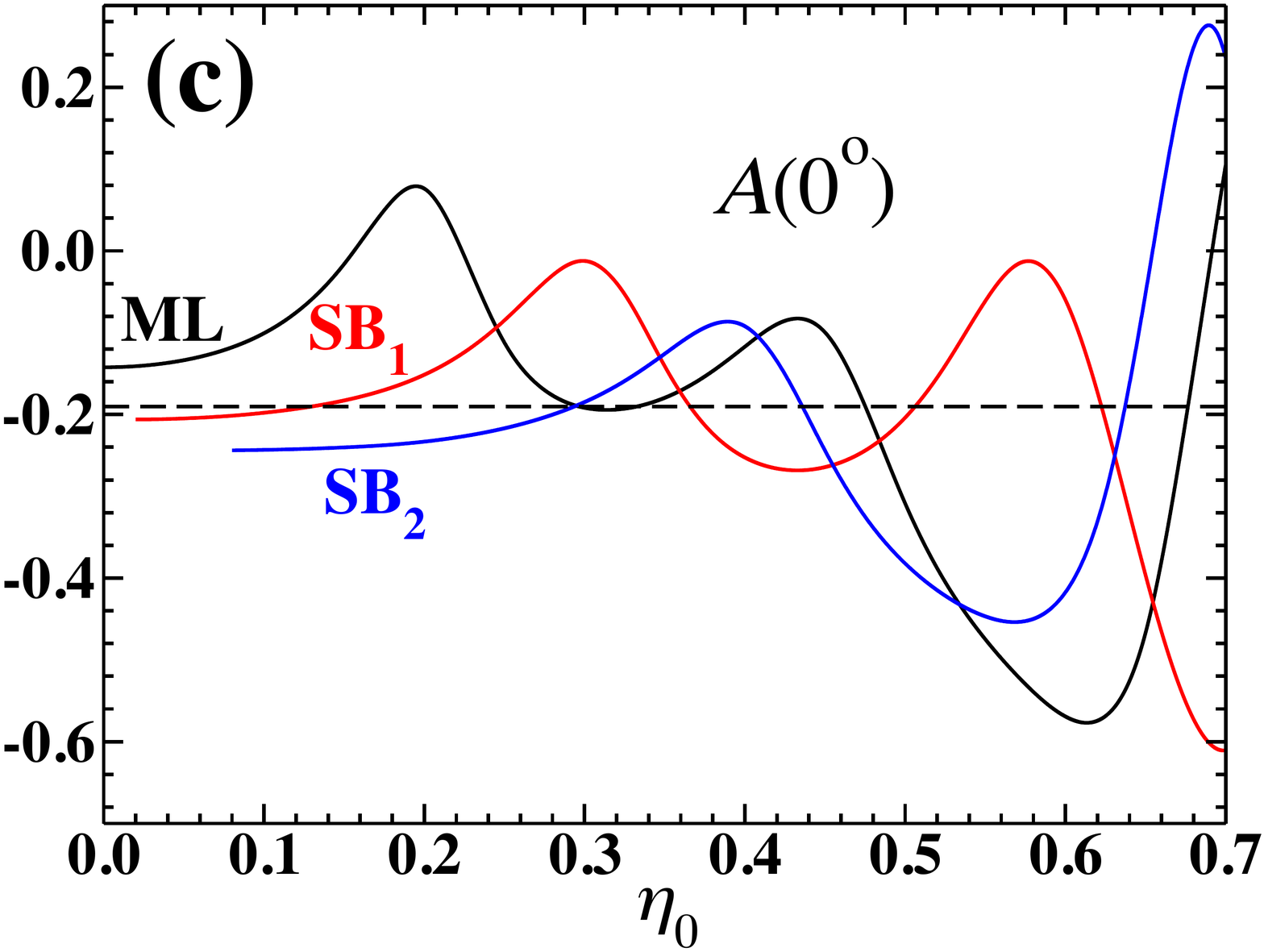}
\end{array}$
   \end{center}
 \caption{Subtracted~(a) and summed~(b) forward and backward ionization probability signals (in units of 10$^{-4}$), 
 as well as the left-right asymmetry~(c), 
 as a function of the IR field amplitude ratio $\eta_0$, as calculated in the TDSE and SFA approaches.
 Note the constant value of~$A(0^{\circ})$ in the SFA. }  
  \label{fig:6}
\end{figure*}

\subsection{Angular Distribution of Sidebands}

Calculations of the anisotropy parameters $\beta_k$ for $k\leq6$ as function
of $\eta_0$ in the TDSE and SFA approaches
are depicted in Fig.~\ref{fig:5}. A first prominent feature consists of
non\-zero values of odd-rank $\beta_k$, which are the result of interference between ionization 
paths involving even and odd numbers of photons. Interestingly, for weak IR fields,
the absolute values of odd-rank $\beta_k$ systematically increase with increasing~$n$.
The ML anisotropy parameters
$\beta_5$ and $\beta_6$ vanish at $\eta_0 = 0$, because only one- and two-photon absorption 
can occur at the ML in the absence of an IR field.
Note also that $\beta_4$ associated with ML is zero
without an IR field
in a simplified version of the second-order PT
with only one resonant $3s$ intermediate state.
Since SB$_1$ and SB$_2$ are, respectively, formed from absorption of at least one XUV and one or two additional IR photons, 
they exhibit non\-vanishing values of~$\beta_5$ and~$\beta_6$ as long as the sidebands can be seen
at small~$\eta_0$.
The good agreement between TDSE and SFA at ML for $\eta_0=0$ is not surprising
since the parameters in Eq.~(\ref{eq:ang_distr}) were extracted from the TDSE code in the absence of IR field.
The small remaining discrepancies, apparent for $\beta_1$ and $\beta_3$, are due to the fact that the PT only includes $s$, $p$, and $d$-waves, 
whereas more partial waves are included in the TDSE calculations.
%Nevertheless, it is remarkable that the SFA predicts the values of the anisotropy
%parameters to such good accuracy at SB$_1$ and SB$_2$ for $\eta_0\ll1$, and particularly well in the case of 
%even-rank anisotropy parameters.

The results from both calculations, TDSE and SFA, clearly exhibit oscillations, although their amplitudes are significantly 
larger in the SFA approach. Nevertheless, the
positions of the maxima and minima, the sign of curvatures, and the limiting values of the anisotropy parameters 
for $\eta_0\ll1$ are in correspondence in the two approaches. 
These oscillations should be experimentally observable. 
Similar oscillations in $\beta^{(1)}_2$ and $\beta^{(1)}_4$ were obtained within 
SFA~\cite{Mazza16} for ionization of helium by circularly polarized XUV and IR radiation.
Such a complicated intensity dependence is determined by the interplay of Bessel functions, 
weighted by the Legendre polynomials, and the denominator in expression~(\ref{eq:beta_general}).
Within the TDSE approach, the oscillations are due to the
contribution of many partial waves originating from
absorption and emission of multiple photons.

The origin of the discrepancies between the TDSE and SFA results can be attributed to several factors.
First, the SFA approach neglects the dynamical atomic polarizability, which becomes 
important at strong IR fields and should perturb
the two-photon pathway through the AC-Stark shifted $3s$~state. 
The disparities could also be due from the SFA assumption that the field is stationary, whereas the pulse
used in the TDSE calculations is still relatively short with a sine-squared envelope. 

It is interesting to evaluate the variation of the left-right asymmetry $A(0^{\circ})$, Eq.~(\ref{eq:asymmetry1}), 
as a function of the IR intensity, since it is an indicator for interference between the amplitudes of odd-photon and even-photon
ionization.
Figures~\ref{fig:6}(a,b) compare the subtracted and summed right ($\theta = 0$)
and left ($\theta = \pi$) emission probabilities calculated in the TDSE and SFA approaches. 
To obtain converged results at large IR intensity, we included
anisotropy parameters $\beta_{k}$ up to $k\le24$.
The TDSE and SFA results agree reasonably well over the intensity range covered.
On the other hand, the asymmetries (Fig.~\ref{fig:6}(c)), i.e., the relative difference between these results,
behave differently in each approach as function of the IR field amplitude ratio~$\eta_0$.
In the SFA, the asymmetry is predicted to have a constant value independent 
of both the band and the IR intensity. This feature is inherent to the SFA and
follows directly from Eq.~(\ref{eq:and1}); the electron in the IR field carries any initial 
asymmetry created in the ML ($n=0$) by the $\omega+2\,\omega$ process over to the different sidebands.
On the contrary, the TDSE results exhibit oscillations of the asymmetry whose amplitude increases 
as a function of the relative IR intensity. 

\begin{figure}[b]
\includegraphics[width=8.5cm]{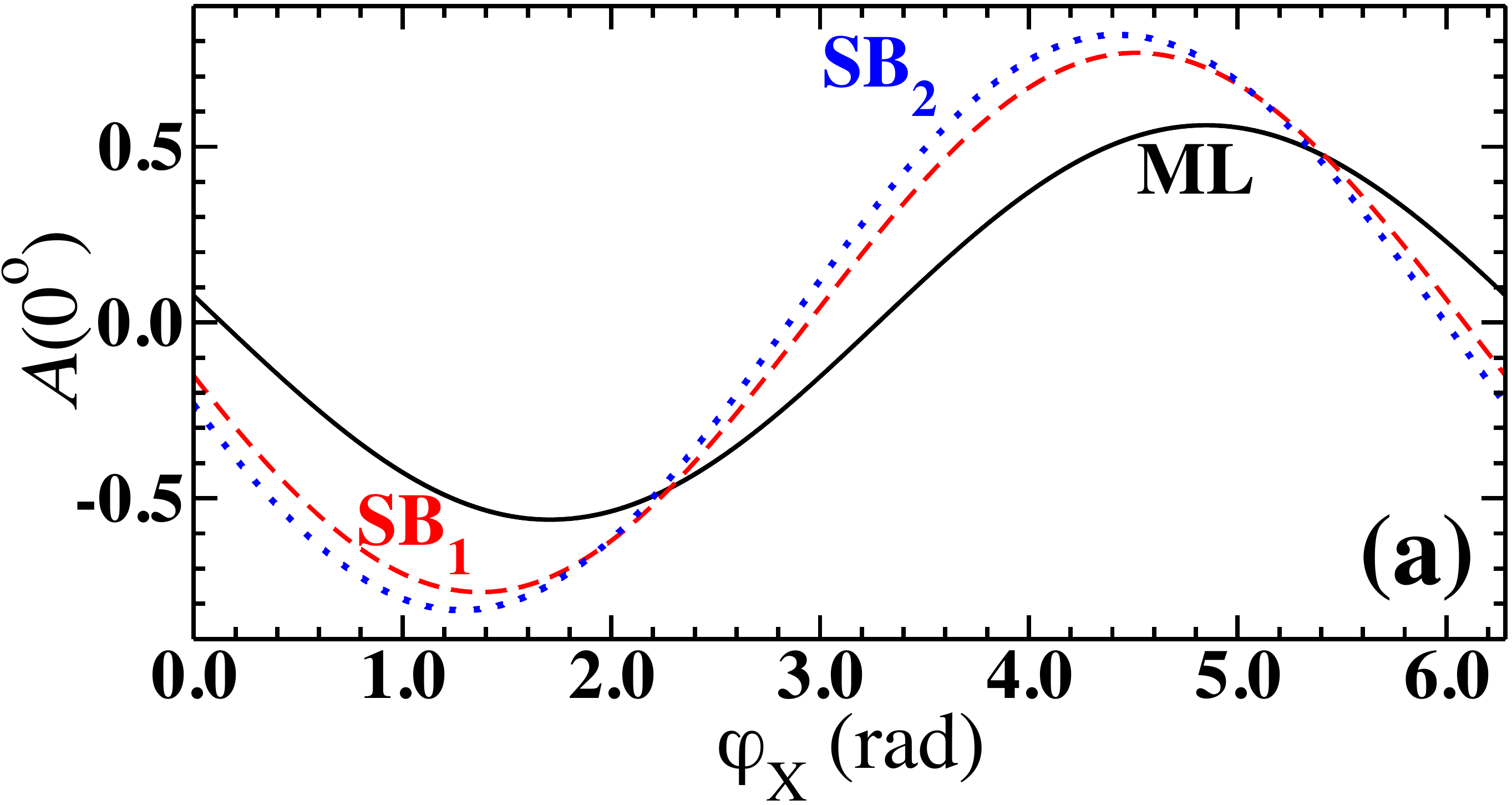}\\
\includegraphics[width=8.5cm]{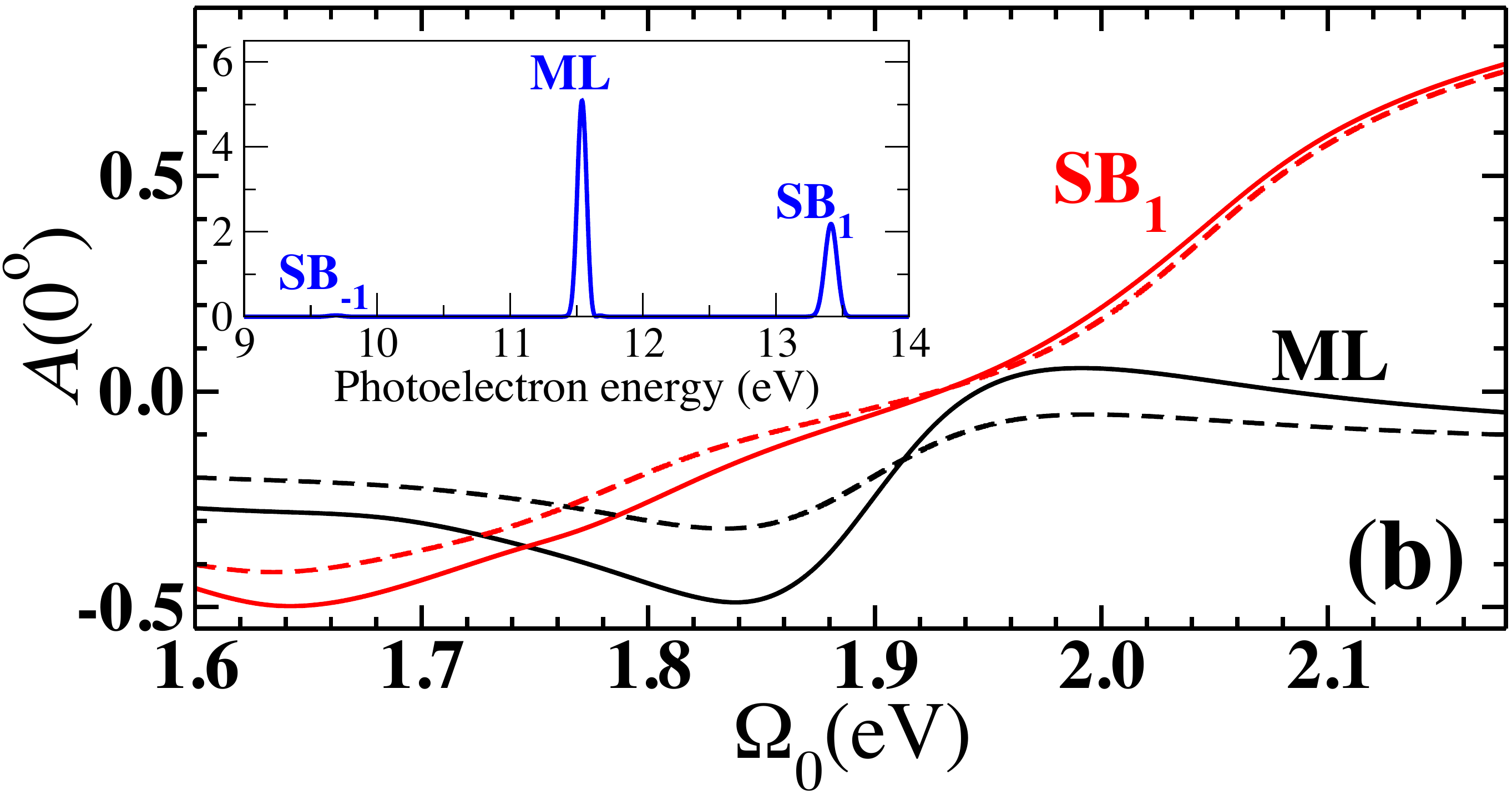}
  \caption{(a) Asymmetry for ML (black solid line), SB$_1$ (red dashed line) and SB$_2$ (blue dotted line), 
  as a function of $\varphi_X$ at $\Omega_0=0.55$~eV and $\eta_0=0.2$. 
  (b)~Asymmetry as a function of the IR frequency $\Omega_0$ at $\eta_0=0.2$ 
  (dashed lines) and $\eta_0=0.3$ (solid lines). 
  The inset shows the corresponding ionization probability spectrum
  (in units of 10$^{-3}$~eV$^{-1}$) at $\eta_0=0.3$.}  
  \label{fig:7}
\end{figure}

Additional information on the asymmetry can be obtained by analyzing its variation 
as a function of the relative CEP between the harmonics.
The left-right asymmetry $A_n(\eta_0,\varphi_X)=a_n(\eta_0)\cos[\varphi_X+\varphi_n(\eta_0)]$ 
at each band has a sinusoidal form,
where (at fixed $\eta_X$) the functions $a_n(\eta_0)$ and $\varphi_n(\eta_0)$ depend on the 
IR intensity only. The variation of
the asymmetry as a function of the relative CEP at $\eta_0=0.2$ is 
presented in Fig.~\ref{fig:7}(a). Although the amplitude of
the oscillations is about 30-40$\%$ larger for SB$_1$ and SB$_2$ 
than for ML, the phase-shift $\varphi_n\le0.5$ rad remains small.
Consequently, the asymmetry at the different sidebands depicts close-lying 
values at all values of $\varphi_X$. Experimentally, this means
that an independent control of the asymmetry at the different bands as a function of 
the relative CEP and IR intensity might only be possible
for sufficiently short pulses, while only a common control of all bands can be achieved for long pulses.

\subsection{Coherent Control of the Sidebands}

The asymmetry is known to strongly vary in situations when one of the 
pathways involves an intermediate resonance. Varying the XUV fundamental frequency 
near the intermediate $3s$ state leads to a Fano-like profile of the asymmetry 
at ML~\cite{Douguet16b,Grum15}, which would be transferred to the sidebands by
a sufficiently long IR pulse. 
To produce a selected control of the bands' asymmetry, an additional resonant path 
with non\-equivalent effects on the different bands should be created. Such a scheme
could be realized by creating an ionizing path different from 
$\omega+2\,\omega$, for example by tuning the IR frequency near the $3s\to3p$
transition ($\Omega_0=1.88$~eV in the SAE model),
as shown in Fig \ref{fig:1}. The PAD associated with SB$_1$ is then expected 
to strongly reflect the opening of the new
pathway. Since our SFA model cannot take into account the $3s\to3p$ transition, 
we only present TDSE predictions below.

Results for the asymmetry at ML and SB$_1 $ as a function of the IR frequency
are presented for $\eta_0=0.2$ and $\eta_0=0.3$ in Fig.~\ref{fig:7}(b). 
The onset in that panel shows the spectrum at $\Omega_0=1.88$~eV for an IR field 
covering $N_0=30$ cycles to overlap with the XUV pulse at this frequency.
Since one can barely see SB$_{-1}$ and only the signals at ML and SB$_1$ are visible for the 
chosen pulse parameters, we present the asymmetry at just ML and~SB$_1$. 

The asymmetry at SB$_1$ exhibits a broad and steady increase, starting from about $-0.45$ 
at $\Omega_0=1.60$~eV and ending at almost $0.80$ at $\Omega_0=2.20$~eV. 
As seen in Fig.~\ref{fig:7}(b), this variation depends only weakly on the IR intensity, 
as opposed to the asymmetry at ML, which clearly shows an asymmetric resonance profile 
whose amplitude increases as a function of the IR intensity. The cause 
of the resonance profile at ML is actually indirect: as the IR frequency~$\Omega_0$
approaches the resonance, a transfer of population from $3s\to3p$ occurs, thereby 
decreasing the two-photon ionization pathway and ultimately
modifying the value of the asymmetry. Consequently, one can control the asymmetry 
amplitudes by varying the IR intensity and frequency. 
The latter effect could, for instance, be exploited experimentally to control the resonance profile 
without having to vary the amplitude ratio between the XUV harmonics, which is a difficult task
to achieve in practice. Nevertheless, a complication arises in the present situation
due to the fact that the system only needs to absorb two IR photons
to ionize from the $3p$ excited state of neon.
As a result, near-threshold ionization might become important for large IR intensities and
could hinder good statistics of the experimental data. 
Thus, it seems preferable to use this scheme on a more strongly bound 
electronic state, for which multiphoton ionization would remain negligible. 

Finally, note that for large IR frequencies in Fig.~\ref{fig:7}(b), the 
asymmetries in ML and SB$_1$ differ significantly. 
It might then be possible, by appropriately varying~$\varphi_X$, to create 
a situation when the asymmetries of ML and SB$_1$ have opposite signs, 
i.e., a situation in which electrons of two different lines have 
opposite preferred emission directions with a resolvable energy difference. 
This strong difference in the asymmetry between ML and SB$_1$
contradicts the SFA prediction. The difference is most probably due to the
$\hbar\omega+\hbar\Omega_0+\hbar\omega$ and $\hbar\omega+2\hbar\Omega_0+\hbar\omega-\hbar\Omega_0$ 
pathways to SB$_1$, which cannot be adequately described in the SFA approach.

\section{Conclusion}
\label{sec:4}
We have presented a detailed investigation of above-threshold 
ionization effects induced by an infrared field on 
two-pathway interference between a non\-resonant one-photon and 
resonant two-photon ionization of neon.
The characteristics of the sidebands in the photo\-electron 
spectrum were analyzed at several infrared laser intensities 
by numerically solving the time-dependent Schr\"odinger equation. 
The numerical results were qualitatively supported by  
analytical formulas derived from a model based on the strong-field approximation.

The ionization probability and the anisotropy parameters 
characterizing the angular distribution at each band agree well with each other in
both models. The anisotropy parameters of the photo\-electron angular distribution 
exhibit oscillations, which should be measurable experimentally. 

%The mainline ionization probability decreases 
%gradually with the infrared laser intensity,
%whereas ionization at the sidebands reaches a maximum before diminishing towards zero. 

An important result of our study is that for long pulses 
the left-right asymmetry at each sideband should not depart
strongly from the asymmetry created at the main photo\-electron 
line in the absence of infrared field. As the infrared intensity
is increased, variations of the asymmetry are shown to become 
significant for relatively strong fields, where the ionization 
signal at each band should be harder to detect. In fact, in 
the strong-field approximation, the asymmetry is 
predicted to be constant, independent of the sideband order and 
the infrared pulse intensity, i.e., the infrared field transfers 
the asymmetry of the main photo\-electron line to the different sidebands.

An interesting situation may occur if the infrared frequency is 
set in resonance with a nearby optically allowed electronic state.
In such a case, the infrared field can play an active role in 
the ionization process. We have shown that the pumping of population 
from $3s$ to $3p$ results in a resonance profile of the asymmetry 
at the mainline as a function of the infrared frequency. 
Furthermore, the amplitude of the resonance profile at the main line
increases with growing infrared intensity. 
The asymmetry of the lowest high-energy band exhibits the most interesting 
variation, since it changes significantly
with the infrared frequency in a monotonic fashion. On the other hand, 
it merely varies as a function of the infrared intensity. 
Initial experiments to reveal some of the features provided by the 
infrared field were already performed at the FERMI 
free-electron laser facility in Trieste (Italy), and future 
experiments could use such effects to improve control
of the photo\-electron angular distribution.

\section*{Acknowledgments}
The authors benefited greatly from stimulating discussions
with Giuseppe Sansone, Michael Meyer, Elena Gryzlova, and Nikolay Kabachnik.  
The work of N.D.\ and K.B.~was supported by the United States National 
\hbox{Science} Foundation under grant \hbox{No.~PHY-1430245} and the XSEDE allocation \hbox{PHY-090031}. 
The calculations were performed on SuperMIC at the Center for Computation \& Technology at Louisiana State University.

\end{document}